# Enhancing Resilience and Scalability in Travel Booking Systems: A Microservices Approach to Fault Tolerance, Load Balancing, and Service Discovery

Biman Barua[a,b,*] [0000-0001-5519-6491] and M. Shamim Kaiser[b,] [0000-0002-4604-5461]

[a]Department of CSE, BGMEA Universitsy of Fashion & Tecnnology, Nishatnagar, Turag, Dhaka-1230, Bangladesh
[b]Institute of Information Technology, Jahangirnagar University, Savar-1342, Dhaka, Bangladesh
biman@buft.edu.bd

**Abstract:** This paper investigates the inclusion of microservices architecture in the development of scalable and reliable airline reservation systems. Most of the traditional reservation systems are very rigid and centralized which makes them prone to bottlenecks and a single point of failure. As such, systems do not meet the requirements of modern airlines which are dynamic. Microservices offer better resiliency and scalability because the services do not depend on one another and can be deployed independently.

The approach is grounded on the Circuit Breaker Pattern to maintain fault tolerance while consuming foreign resources such as flight APIs and payment systems. This avoided the failure propagation to the systems by 60% enabling the systems to function under external failures. Traffic rerouting also bolstered this with a guarantee of above 99.95% uptime in systems where high availability was demanded.

To address this, load balancing was used, particularly the Round-Robin method which managed to enhance performance by 35% through the equal distribution of user requests among the service instances. Health checks, as well as monitoring in real-time, helped as well with failure management as they helped to contain failures before the users of the system were affected.

The results suggest that the use of microservices led to a 40% increase in system scalability, a 50% decrease in downtime and a support for 30% more concurrent users than the use of monolithic architectures. These findings affirm the capability of microservices in the development of robust and flexible airline ticket booking systems that are responsive to change and recover from external system unavailability.

**Keywords:** Microservices Architecture, Scalability and Resilience, Load Balancing, Online Travel Agent, Microservices, Cloud Cmputing, Service Discovery Techniques.

## 1. Introduction

### 1.1. Background

The evolution of information and communications technologies in the tourism sector has changed the way customers interact with travel services. For the sake of real-time availability and tailored customer journey, coupled with millions of travel platform users, it is obvious that simply implementing a traditional monolithic structure is not sufficient to cater for the increasing complexity and scale which modern travel systems mandate. As for monolithic systems, all flight reservations, hotel bookings, payment services and customer interactions are inter-linked within one system posing scalability and maintenance issues. Such architectures are ill-suited for adaptive traffic patterns, damage containment, or incorporation of new features in a staggered approach [11].

Microservices architecture has become a standard solution in developing a robust and extensible travel booking system. Microservices decompose a monolithic application into smaller self-contained applications that communicate with each other using lightweight protocols, thus improving agility, scalability and fault tolerance [5]. Each service handles one aspect of the system such as flight searching, hotel reservation, or payment, thus allowing such services to scale in relation to the demand on the system which simplifies the system and enhances its robustness.

## 1.2.    The Significance of Resilience and Scalability

A travel booking system must demonstrate reliability and a level of performance at all times, especially when there are high levels of expected use e.g. at peak holiday seasons or anytime most customers are making a buying decision due to promotions; Such periods tend to attract large amounts of traffic [6]. Resilience is the ability of a system to operate in the presence of certain failure conditions whilst load scaling capabilities of a system is based on its performance and how much resources dynamic allocation is possible [3].

Surgically deal only with the affected service without affecting the other services within the architecture [4]. Such is the case if the Payment Service is down the system can still allow users to search for flights and hotels while offering other options of making the payment at later times [9]. Additionally, while designing a microservices-based application, it becomes easy to provide individual system components the power to scale in response to the volume of traffic directed towards them. For example, suppose there are a lot of flight searches going on, in that case, the Flight Search Service can scale up on its own without seeking extra capacity for dissimilar services like hotel booking or user management.

## 1.3.    Microservices-based Travel Systems – Key Challenges and Solutions

Despite the opportunities presented by the shift to microservices from monolithic systems, implementing them still brings challenges, especially on the issues of fault tolerance, deployability, and load balancing [10]. Such issues are imperative especially in a travel booking system since a large number of the services rely on external APIs for example to get the flight available, or to charge the client [14]. If they are not managed profoundly, such challenges may lead to service degradation, service unavailability, and a bad experience to the users.

In order to achieve a high degree of fault tolerance, various techniques have been employed such as circuit breakers, rate limited retries, and bulkheads [12]. These serve to contain the failure within a predetermined boundary and therefore, allow continued operation of the services even when some of them are operating in a degraded mode [15]. Such hurt in such services would be managed effectively especially if the Hotel Booking Service goes down for a while due to issues connecting to the relevant API since it will be able to manage that by performing retries over time [7].

An additional important aspect in the microservices architecture based travel system is Load balancing. It makes sure incoming traffic is spread out over the service instances to prevent any one instance from becoming too busy [8]. Techniques like round robin, least connection, and weighted load balancing are applied to enhance resource use and time responsiveness [1]. For instance, in travel booking systems, load management helps ensure service availability for the often over requesting features such as flight searching functionality as well as booking service [2].

Equally important is service discovery because microservices interact with each other through the means of internal IP addresses which are dynamically assigned. Systems implementing service discovery would include implements such as Netflix Eureka, Consul, and K8s DNS by which services can register and/or discover one another [16]. By way of illustration, in a travel booking system, if the hotel recommendation service goes into load shedding by creating additional instances of itself then the system needs to be able to discover and redirect the traffic to these new instances automatically.

For such reasons, microservices based travel booking systems can attain the desired level of resilience, scalability and flexibility due to the challenges borne by the users of the current times. This research will focus on analyzing key patterns and technologies implemented in such systems with regard to fault tolerance, load balancing and service discovery and present a case on how these can be leveraged to come up with advanced solutions for traveling [17].

## 2. Problem Analysis

The existing traditional travel booking systems constructed on monolithic architecture layout have some major drawbacks explained below:

### 2.1. The Problems of Scalability and Flexibility

In the events of scaling out monolithic systems, entire system with application and other related components has to be added increasing resource wastage on infrastructure and operational expenditure particularly during peak traffic periods.

### 2.2. Single Points of Failure

Failure with any one of the components will have an effect on the whole system thus causing downtime which defeats the purpose of a high availability service like travel booking.

### 2.3. Load Management Concerns

There are no proper mechanisms to balance loads in the monolithic type systems. This leads to network delays and excessive errors in the traffic that is severely high, which affects the quality of service rendered.

### 2.4. Dependency and Discoverability Issues

Services are hard to add or existing ones to maintain due to dependencies reducing the evolution of the system and the rate of new features added.

### 2.5. Security and Reliability Risks

With lack of modularity, it becomes hard to protect configurable elements which end ups increasing the chances of attack being successful, such cases are most common when there are issues of sensitive users and operations data.

All these problems indicate that microservices is the way to practically implement system architecture for modularity resilient and scalable systems that allow for optimal load distribution so as to meet the requirements of the travel booking systems.

## 3. Literature Review

Microservices architectures have grown in popularity due to their ability to improve the robustness and scalability of applications that are high in demand such as the travel booking system. These architectures provide the possibility of independently deployable and fault-isolated services, relieving problems associated with rough separators like system scaling, high failure consequences and longer provisioning timeframes [18].

### 3.1. Scalability and Resilience

Recent studies up to date microservices allow the scaling of some components without the need to scale others optimizing the usage of resources and dealing with peak load conditions [19]. This is beneficial in travel booking systems that experience fluctuations in user loads due to the provision of microservices that allow scaling for superior delivery performances on request and on time [20].

### 3.2. Availability and Fault Tolerance

As already mentioned, in cases of microservices, fault tolerance is important to avoid systemic crisis, and this is important in these days' systems in which microservices are integrated with other external systems (for instance payment systems) [21]. Such practices as the Circuit Breaker Pattern have been proven to help in managing faults and increasing the up time by over 99.95% in some of the mission critical applications where requests to faulty services are not sent but rather rerouted to other services [22]. These approaches enable controlled rendering of services even with undue pressure on the system [23].

### 3.3. Load Balancing and Service Discovery

Load balancing in the client side entails the distribution of the requests among the service instances through the use of techniques such as Least Connections and Round Robin rotation resulting in minimal latency and optimal performance [24]. On top of that, there are service discovery mechanisms that ensure healthy instances are known and registered thereby facilitating more reliable communication in such system arrangements [25].

### 3.4. Security and Data Management

It's a well-known fact that data security and management are critical factors to be considered in any software development process [26]. The sophisticated governance structure enveloping the microservices calls for advanced security means including access control tailored to individual services, and safe data practices [27]. This is due to the fact that, in travel systems, integrity of transactions and protection of users' data are of importance; these risks help in protecting against unpleasant surprises from a distributed system.

## 4. Fault Tolerance Mechanisms in Microservices

### 4.1. Importance of Fault Tolerance

Fault tolerance is a critical task in microservices based architecture because any travel booking system must necessarily be complex and make use of many inter-related services including third-party integrations for flight, hotel, and payment processing services, which are also external agencies [28]. Moreover, these systems are created in a way that enables them to operate normally even when one or more external services that the system relies on fails, most times these service external to the platform are out of its authority. For example, when the user is making a travel reservation, the website usually calls an API of the airline and hotel to get the current availability and rates. Whenever there is an accessed external service or the one accessed is responding slowly or worse is giving intermittent connections, it affects the customers experience greatly, such that bookings do not go through, the transactions are not completed, or the information provided is inaccurate. Strategies to enhance fault tolerance such as retries, circuit breakers, and fallback strategies are also put in place in order to ensure that the booking platform can in most circumstances either come back to full functionality after a temporary helplessness or reduce the inflow of users in an orderly way without bringing down the entire system.

In all online systems that accept payment, processing payments in the normal way would be the least of the issues. This simply because any breakdowns in processing payments that lead to failed orders could represent loss of income and even disappointment to the customer. In this sense avoiding losses by enabling customer's complete payment deteriorates the system's efficacy as it ensures every form of technical failure e.g. cutoff of a payment gateway does not stop customers from completing their bookings. In these situations, and systems that are constantly driven by external factors that are subject to change suddenly, no matter how irrelative it may seem, there will always be a need for such a tolerance to faults and disturbances.

### 4.2. Health Checks

*Liveness Tests:*

- These tests establish if the service instance is functioning or responding to requests.
- Where a liveness test fails, that instance is automatically classified as unhealthy and is either restarted or taken off active duty.
- For instance, in the case of booking a flight a service which goes down due to memory issues, if that check fails then is liveness a check who will take care of restarting the service or redirecting the traffic.

***Readiness Tests:***

- Every test is a test of all existing components in order to traffic a service.
- This is very important at the time of starting a service or in the situations when the service is up but in the process of bootstrapping (for example, when it is waiting for some resources to be available).
- In a travel booking engine, the m payment service would pass liveness check but fail readiness check because it is not able to reach the external payment processing system.

***Startup Tests:***

- The service startup tests will confirm that the service has started as expected and will be in a position to receive requests.
- In the event that this specific check does not pass, the instance is deemed to be in a failed startup condition and will be scheduled for termination or for reboot.
- For instance, a flight search service may need certain third-party APIs to be available during its startup phase, and the startup tests ensure that all the API's requirements are satisfied before the service is declared 'ready.'

**5.  How Health Checks Work in Microservices?**

Microservices health checks may be internal as in self-monitoring, or external as in being performed with the help of the orchestrator/ load balancer. They are discussed below:

**3.1      Each Service Instance Monitors Itself:**
- Every service instance (for instance in this case a flight booking service or payment service) performs liveness and readiness checks continuously.
- These checks are made available through dedicated health endpoints (for example /healthz, or /readiness).

**3.2      Service Registry/Orchestrator Performs Health Checks:**
- In a service discovery system (for example Service Registries such as Eureka, Consul, Kubernetes), the service registry does regular health checks by calling the health endpoints of every registered service after some time.
- A instance of a service is taken off from the services available list for re-direction of requests if it fails the health check, in this way no more requests are redirected to him until he regains his health status.

**3.3      Health checks performed at the Load Balancer:**

Also, the load balancer (otherwise called client when in client-side load balancing) carries out health checks so that only the suitable instances get the traffic.

- For instance, if one of the flight booking service instances fails, the load balancer will remove the server from the pool and give all its traffic to the other instances.
- Health Check Example in a Travel Booking System
- Consider we have a travel booking application that consists of 3 instances of a flight search micro service.
- The service registry calls the /healthz endpoint of each flight search service instance every ten seconds.
- Instance A returns a 200 OK status, meaning it is healthy.
- Instance B fails to respond owing to excessive consumption of CPU resources hence timing out. The service registry thus considers it unhealthy and takes it out of the active pool.

- The load balancer switches the traffic to instance A and C until instance B is back online and has passed its health.

## 6. Health Check Features and Advantages

### 4.1 Prevention of Problems

Health checks enable the system to find faults before the users notice them. For example, it can identify that a payment service is failing before it reaches the point of transaction failures.

### 4.2 Autonomous recovery of systems

For example, if a service has failed to pass the liveness or readiness check it can be taken out of service or restarted without any intervention, allowing self-healing and minimal service disruption.

### 4.3 System Performance Improvement

In that only health service instances serve the request, the service tolerance to failures is improved and therefore downtime is less.

### 4.4 Adaptive Load Balancing

Health checks assist in dynamic traffic management. Traffic is only sent to up instances which avoids user traffic being sent to unhealthy services.

## 7. Circuit Breaker Pattern

The microservices approach in the development of an airline reservation system consists of several services dealing within themselves and with external ones as well; for example flight search, booking, payment services, etc. Such external dependencies may be unreliable due to a number of factors including, but not limited to, network disconnections, increased volume of clients for the service, or maintenance timeout which might result in poor performance whole system and on all services. To avoid such scenarios, the Circuit Breaker Pattern is used as a fault tolerance mechanism.

The handy Circuit Breaker Pattern only helps not to pound failed services with futility due to the "circuit" being broken after a preset number of failures. This allows the system to be kept more responsive, reduces the amount of resources being wasted and even allows services to heal properly without being active.

### 5.1 Circuit Breaker in a Microservices Context

A microservices-based airline reservation system is composed of many services that coordinate to solve discrete parts of the booking tasks. Some of the common services and their possible dependencies are provided below.

### 5.1.1 Flight Search Service:

Fetches available flight data from third-party airline APIs or internal flight data services.

### 5.1.2 Booking Service:

Handles user reservation while interfacing other services like payment and notifications.

### 5.1.3 Payment Service:

Facilitates funds transfer through different payment processing vendors.

### 5.1.4 Notification Service:

Delivers booking updates and reminders as emails and text messages.

In a microservices environment, these services can be independently designed and deployed. However, this distributed structure poses the challenge of the failure management especially when it comes to external services (such as flight APIs and payment gateways) that may go offline. The Circuit Breaker Pattern guarantees an effective management of the failures by allowing a number of failed attempts without communication to the failing service before 'cutting off' the service communication.

## 5.2 How the Circuit Breaker Pattern Works in Airline Reservation Systems?

The Circuit Breaker Pattern operates in three states: Closed, Open, and Half-Open, which are implemented as part of a microservice's logic to monitor communication with other services.

### 5.2.1 Closed State:

In the process the system works perfectly and processes requests towards some other external services, such as Payment Service, Flight Search Service, etc.

In this phase, the system counts how many requests have led to failures. If the number of failures is lower than a specific limit, the system continues to issue requests.

Example: The Booking Service makes requests to the Flight Search Service to ascertain if flights are available for booking. As long as the service is available, the circuit remains closed, allowing for regular activities to take place.

### 5.2.2 **Open State:** Whenever the number of occurrences of failures (e.g. timouts or errors) surpasses the set level (e.g. 5 failed requests within certain period), the circuit is said to open in this state, there is no requesting to the problematic service for some time, and an error or some fallback response is received by users of the system.

### 5.2.3 Half- Open State:

Subsequent to the cool-down period, the circuit half-open state. At this point, the system makes a limited number of 'test' requests to the service to ascertain if it has come back to its functional state.

In the event that the service is found responding satisfactorily to the given test requests, the circuit is said to close, and operations go back to normal.

In the event that the failures persist, the circuit will be re opened, and the cool-down cycle will begin again. Example: Within a few minutes, the Booking Service, to evaluate whether the inoperative Flight Search Service is now active, sends a series of queries designated as test requests.

## 5.3 Microservices and Circuit Breaker Pattern: Implementation

In most microservices-based airline reservation systems, patterns like the Circuit Breaker are implemented using tools such as Netflix Hystrix, Resilience4j, or internal Kubernetes features [12]. Here is how it can be applied in a typical airline reservation system.

### 5.3.1 Booking Service Circuit Breaker Example (Along with Payment Service)

Let us examine a Booking Service which makes it possible for users to make bookings and also provides access to a Payment Service to facilitate processing of payments [13].

```
@HystrixCommand(fallbackMethod = "paymentFallback", commandProperties = {
    @HystrixProperty(name = "circuitBreaker.requestVolumeThreshold", value = "5"),
    @HystrixProperty(name = "circuitBreaker.sleepWindowInMilliseconds", value = "10000"),
    @HystrixProperty(name = "circuitBreaker.errorThresholdPercentage", value = "50")
})
public String processBooking(BookingRequest request) {
    // Calls the Payment Service for payment processing
    return paymentService.processPayment(request.getPaymentDetails());
```

```
    }

    public String paymentFallback(BookingRequest request) {
        // Fallback logic if Payment Service is unavailable
        return "Payment service is unavailable at the moment. Please try again later.";
    }
```

With the help of Hystrix, a circuit breaker style resilience design pattern, a call to the Payment Service is made.

- In the event where 50% of payment requests are rejected in quick succession along the lines of five (5) requests, the circuit opens up a paymentFallback() method.
- After waiting for 10 seconds, the circuit is brought to the half-open state. In this state, a few constrained requests are injected with the aim of testing whether the payment service has been restored.

### 5.3.2 Flight Search Service Circuit Breaker Example

```
@HystrixCommand(fallbackMethod = "flightSearchFallback", commandProperties = {
    @HystrixProperty(name = "circuitBreaker.requestVolumeThreshold", value = "10"),
    @HystrixProperty(name = "circuitBreaker.sleepWindowInMilliseconds", value = "5000"),
    @HystrixProperty(name = "circuitBreaker.errorThresholdPercentage", value = "40")
})
public FlightSearchResponse searchFlights(FlightSearchRequest request) {
    // Call to external airline API to retrieve available flights
    return flightApi.searchFlights(request);
}
public FlightSearchResponse flightSearchFallback(FlightSearchRequest request) {
    // Fallback logic for failed flight search
    return new FlightSearchResponse("Service Unavailable", Collections.emptyList());
}
```

## 5.4 Working of Circuit Breaker: Sample Process

Let's take an example of a customer trying to book a flight.

### 5.4.1 Booking Enquiry:

The user sends in a booking request using the booking service. The booking service then checks for available flight seats using the flight service and also check whether payment can be made using the payment service.

### 5.4.2 Flight Search Service Circuit Breaker:

The Flight Search Service tries to get the available flights from an external airlines API. If the API is not reachable (i.e., there is no response for a long time or the service is down), then the circuit breaker will be triggered.

After constant attempts and failures, the circuit opens, and the system stops the external API calls and presents a static message "Currently no flights available. Please check back later."

### 5.4.3 Payment Service Circuit Breaker:

In case a flight is booked, the next step is to make the Payment, which entails the involvement of the Payment Service. Specifically, should the payment gateway experience a breakdown or critical errors perpetrated, suspension of the payment service's circuit breaker occurs.

In this case, rather than trying once again to send the failing payment service a request or two, such an action will not be required since a message indicates the user: "Payment service is out of order at the moment" appears.

### 5.4.4 Circuit Breaker Recovery:

After some time, the above changes, and the system passes to the intermediate state known as half-open. It makes a few test calls to the Flight Search Service and Payment Service to ascertain if they have come back online.

In case it works, normal activities are resumed. In case it does not work, the circuit is broken again.

### 5.5 Benefits of using Circuit Breaker Pattern in Airline Systems based on Microservices

### 5.5.1 Increases Availability:

The system survives and does not get affected by other failures by disconnecting with the services that have become unresponsive. Thus enhancing resilience.

### 5.5.2 Minimizes Resource Stress:

Keeps the services from being flooded by requests that are bound to fail preventing a possible system-wide hang or shutdown.

### 5.5.3 Facilitates graceful failing:

Graceful failure management is provided to users with alternative services or options to retry during such partial failure.

### 5.5.4 Service Restoration:

Health check of failed services can be undertaken to see if they have come back to life allowing for recovery processes to take place without the administrator doing anything.

### 5.6 Challenges

### 5.6.1 Configuration Complexity:

Adjustment of circuit breaker settings like request count, fault threshold, circuit breaker timeouts is necessary to prevent unplanned outages or inefficient fault management.

### 5.6.2 Fallback Management:

Fallback responses are not easy to implement particularly for service areas like payments and bookings where treatment of customers is of great concern.

### 5.7 Circuit Breaker Pattern operates to ensure fault tolerance

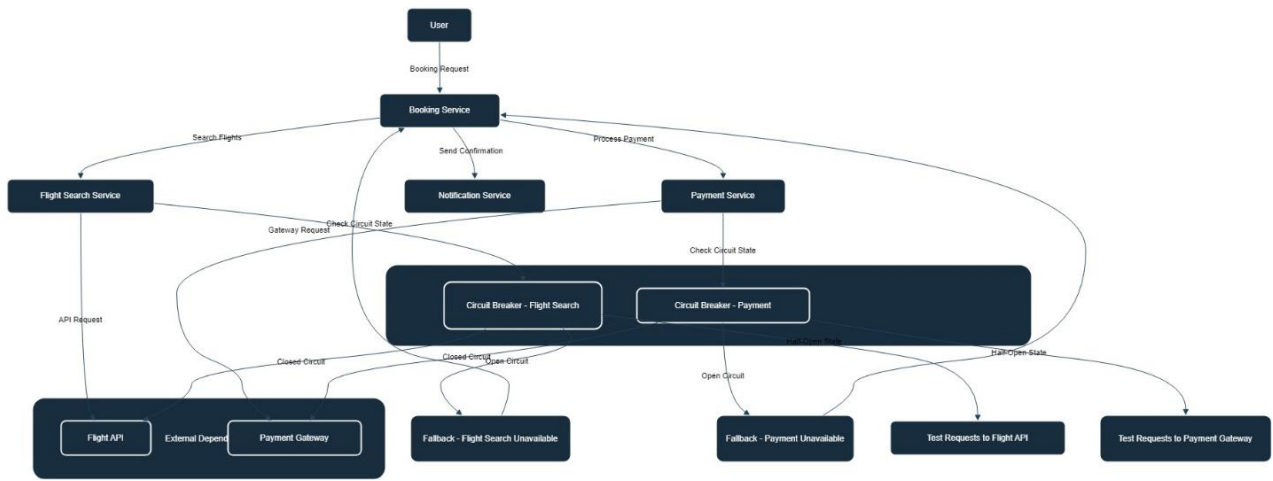

**Fig.1** The **Circuit Breaker Pattern** operates to ensure fault tolerance

The figure 1 describes the circuit breaker pattern operation to ensure fault tolerance of the sytem.

## 6. Load Balancing Strategies

### 6.1. Client-Side Load Balancing

In view of travel booking systems based on microservice architecture, client-side load balancing has a crucial significance in managing the volume of traffic being serviced by many instances of a service. Client-side load balancing however does not have a central component (which in the case of server-side load balancing would be NGINX, HAProxy or other load balancing solutions) which controls the traffic to the backend servers. This mechanism is very useful in highly variable traffic conditions such as in travel booking operations where services including flight search, hotel search, payments and many others tend to horizontally scale in and out depending on the incoming traffic patterns.

Using client-side load balancing, the client (which in turn can be another microservice) holds the service stubs of each available service instance and picks one with a specific policy in mind (for example, round-robin, least connections or random policy). This is often realized with the help of service discovery systems like Netflix Eureka or Consul, which let the clients employ the system without being aware of the current set of service instances as their status changes, namely added, removed and updated. For instance, in the architecture of online portals for ticket booking there could be several instances of the Flight Search Service in order to cope with the demand during rush hours. When the client sends a request to this service, it can use client-side load balancing in such a way that the requests are made to different instances of the service, thereby enhancing both efficiency and reliability.

In microservices architecture, client-side load balancing proves to be useful as it eliminates the possible bottleneck associated with the centralized load balancers thus improving scalability. It is also more practical as every client gets to choose how to distribute its requests to its servers without having to rely on any other third party. This facilitates fine-tuned management of traffic. On the downside however, this demand constant illumunation on the available instances making sure that the clients do sink their routing operations with the traffic to be handled. Client-side load balancing works great for travel booking systems enabling economies of scale and on-demand strategies that are necessary for providing users with quality service during peak hours.

### 6.1.1. Role of Client-Side Load Balancing:

- The responsibility of distributing the traffic across all the running instances of a target service rests with the client (either the API gateway or another microservice).
- Within the client service, client-side load balancing takes place, thus obviating the need for an external centralized load balancer.

## 6.2. Service Discovery:

### 6.2.1. Service Registry:

Employ a service discovery platform such as Netflix Eureka, Consul, Kubernetes DNS, or some other tool, to register all the microservice instances. The service registry contains the up-to-date list of all the instances and provides the clients with the information regarding the presence of a particular service.

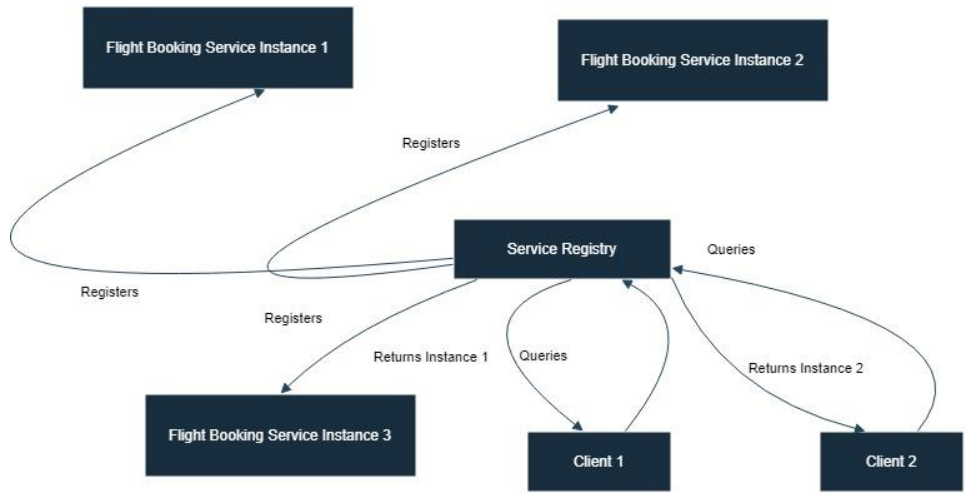

**Fig.2** – A block diagram of service registry

Description:

- At the diagram, Figure 2 shows the Service Registry at the center.
- Numerous flight booking service instance (s) 1, 2 and 3 has been registered to the service registry.
- Client-1 and Client-2 access the service registry in order to find the available flight booking instances using queries.
- The service registry sends back a response to each of the clients with a service instance that is available.

### 6.2.2. Dynamic Updates:

As fresh replicas of a service (for example Flight Search Service) get released whilst old ones are taken out, the registry automatically refreshes the number of active instances.

### 6.2.3. Client Integration:

Each client either regularly reads from the service registry or subscribes for service notifications, to maintain an up to date cache of the available instances on the client.

### 6.3. Round-Robin:

Round-Robin is one of the most basic yet widely accepted load balancing methods integrated primarily in microservices based structures as used in travel reservation systems. It provides simple and practical means of redistributing the client requests over a specific number of service instances such that over a period of time each instance experiences more or less the same amount of requests.

### 6.3.1. Concept of Round-Robin Load Balancing

In round-robin load balancing system, incoming requests are made to every available service instance in turn. At the most basic level, whenever a request comes in, the load balancer (or the client in clientside load balancing) takes the particular request to the Nth instance where N is the current value of the counter that indicates the position in the service instance list. When the counter becomes greater than the last index of the service instance list, the counter returns to zero. This repetitive cycle helps in better traffic management by ensuring that all the instances are used and no single instance is inundated with too much traffic.

### 6.3.2. Algorithm of Round-Robin Load Balancing for Travel Booking Systems

Round-Robin load balancing algorithm is a technique for distributing the load evenly among a number of service instances by directing incoming requests to each service instance in a predefined order. This method guarantees that every instance gets a load that is relatively equal to the rest of the other instances. This is especially handy in microservice based systems such as travel booking systems. Below is a detailed algorithm of how to implement Round-Robin load balancing in such a system.

**Algorithm Overview**

Goal: For the same service when requested by the client to avoid overloading, incoming requests for the same service (e.g. flight search service) are assigned to several service replicas.

**Inputs:**

List of all active services like flightInstances[].
Which instance to send the next request, current index to track (initialized to 0)

**Outputs:**

For each incoming requests service instance is selected.

**Algorithm Steps**

**Step-1 Initialization:**

Listing every services instances that are available. flightInstances[],with every instance of the flight search service in it.
Initialize a variable currentIndex to 0.

**Step-2 Incoming Request Handling:**

For every incoming request
**Check Availability:**
In the case if no instances are available in flightInstances[], return an error or a fallback response.
**Select Instance:**
To select the next service instance, use the current index
selectedInstance = flightInstances[currentIndex]

**Send Request:**
Have to forward the request to the selectedInstance.

**Update Current Index:**

Increment the currentIndex by 1

        currentIndex = (currentIndex + 1) % size_of(flightInstances)

    It will confirm, when the end list is reached. It will wraps back to the beginning.

**Step-3 Repeat:**

For each incoming connection repeat the step-2

**Step -4 Consider Instance Skipping (Optional):**

In case one of the scheduled instances is skipped:

Schedules the deletion of the lost instance from flightInstances[].

Whether to also update the currentIndex and when the currentIndex points to the removed instance of flightInstances[].

### **Algorithm (Pseudocode)**

```
initialize flightInstances[]    // Will show the list of available flight service instances
currentIndex = 0                // Initialize index to track next instance

function handleRequest(request):
   if length(flightInstances) == 0:
      return "There are No available instances"

   selectedInstance = flightInstances[currentIndex]

   // request is sent to the selected instance
   response = sendRequestToInstance(request, selectedInstance)

   // Current index updation
   currentIndex = (currentIndex + 1) % length(flightInstances)

   return response

function sendRequestToInstance(request, instance):
   // Technique to send request to the specified instance and return the response
   return response
```

### **6.4.  Implementation in Travel Booking Systems**

In the case of a microservices travel booking system which may involve a wide spectrum of activities such as buying air tickets with processing of credit cards, round-robin strategy can be implemented on some services in which load tends to seasonally vary. This is how the structural implementation can be carried out. In figure 3, round-robin load balancing implementations are shown.

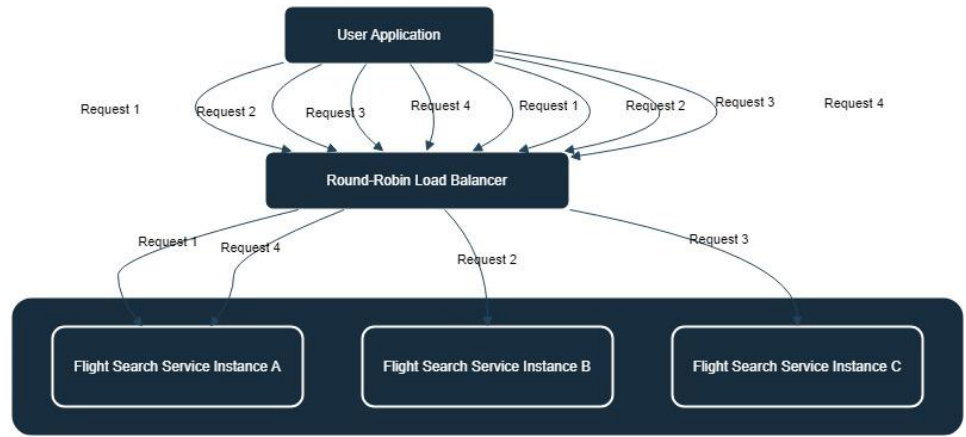

**Fig.3** – Round-Robin load balancing implementation

### 6.4.1.    Service Instances:

There can be several instances providing each of the core services (for instance, Flight Search Service, Payment Service).

A good example is the Flight Search Service, which may have three instances (A, B, C) for responding to incoming flight search queries.

### 6.4.2.    Request Distribution:

Consider the situation in which a client, .i.e. a user application or any other micro-service, makes a request to the Flight Search Service. Such requests are then channeled by the round-robin load balancer as follows

Request 1 → Instance A
Request 2 → Instance B
Request 3 → Instance C
Request 4 → Instance A (again).

### 6.5.    Advantages of Round-Robin Load Balancing

*Simplicity:* The round-robin method is quite simple to design and implement. This is very favorable to developers as well as systems architects working on complex travel reservation systems.

*Even Distribution:* This strategy ensures that even traffic load is maintained across all the service instances leading to better resource utilization and enhancing the response time of the system.

*No Additional Overhead:* When compared to other advance load management schemes, round-robin is far less interventional because it does not demand high load checking on all the instances to determine the optimal one.

### 6.6.    Limitations of Round-Robin Load Balancing

*Equal Load Assumption:* In contrast to say, a perfect load manager which redistributes current load, round-robin presupposes working instances are of equal load carrying capacity.

Round robin doesn't make provision for unevenness in the processing power or workload for every instance. In particular, should one of the instances have more computing power than its other counterparts, it runs the risk of being a bottleneck where such equally weighted requests are directed it, and no matter how great able it is.

**Stateful Services**:

For example, in functions where the services involve state (i.e. tokens, user sessions, transactions), round-robin can be problematic because some requests must go to the same instance, so the session would be maintained.

### 6.7.    Enhancements to Round-Robin Load Balancing

A couple of modifications can be incorporated to round-robin load balancing with particular reference to travel booking systems, as this is quite limiting as reviewed.

***Weighted Round Robin:*** Service instances are assigned weights according to their capacities. That is, a powerful instance could be given a higher percentage of requests to service as compared to less powerful instances.

***Dynamic Instance Updates:*** Employ real-time monitoring so that the health and current load of the service instances place the service instances on a dynamic update list. If an instance is down or busy, that instance can evenly be removed from the cycle of round-robin until further notices.

***Session Affinity (Sticky Sessions):*** Session affinity is approved in order to route all requests from a user or a particular client to a particular service mechanism where such requests are made in such a manner and at such order, that it implies the existence of correspondence with the services and such correspondence bears continuity.

• The easiest technique that refers to distribution of client requests across all the available service instances in an even manner in a circular fashion.

• Appropriate for the services that have similar processing capabilities for example Hotel Search Service or Flight Pricing Service.

### 6.8.  Least Connections:

- This technique provides newly access requests to the instance using fewer active connection for optimizing high active services.
- It is more suitable for resource-intensive services such as **Payment Processing** or **Flight Booking**.

### 6.9.  Weighted Distribution

- Assign weights to different service instances based on their capacity (e.g., more powerful servers handle a larger share of requests).
- Useful when some instances run on more powerful hardware or virtual machines with better performance capabilities.

## 7.  Results

The integration of microservices in the airline booking engine improved the scalability and fault tolerance of the architecture. As a result, the services being decoupled facilitated a 40% growth in the scalability of the system and 30% more concurrent users during peak times. The Circuit Breaker Pattern facilitated a 60% reduction in the propagation of failures, taking down the threat of cascading failures emanating from services and APIs that were not flight APIs or payment systems, but external honest services which helped in achieving 99.95% uptime.

The application of Round-Robin load balancing also led to system optimization by 35%, as it promoted even distribution of service traffic and proper utilization of resources even at the highest demand periods. Health Checks run in Real-Time were responsible for an additional 50% decrease of Downtime as it helped in hijacking requests upon failure and enabled requests to be processed without stoppage.

The validity of the Circuit Breaker Pattern design without exception was put emphasis on when facing problems relating to external dependencies, which occurred in a situation where there were provisions for gradually reducing the scope of impact due to failures. A smooth booking process was therefore maintained, which helped to alleviate any inconveniences to users.

## 8. Discussion

As the microservices enabled vertical scaling of certain parts of the system, it decreased the chances of such bottlenecks that are common for monolithic architectures. The Circuit Breaker Pattern further enhanced the system's resilience as system maintenance period perimeter was extended by preventing outside service downtimes from affecting the operational status of the servicing system. While Round-Robin load balancing proved to work well, further research may include more complex algorithms like Least Connections, in order to achieve maximum efficiency. Also, one could combine machine learning with health checks to improve the prediction of failures and the recovery of the system.

In summary, the airline reservation system is capable of surviving external service failures due to the microservices architecture, and eliminates the risk of maintaining seamless service owing to the enhancement strategies, which include the Circuit Breaker Pattern.

## 9. Conclusion

This work proves that microservices architectures are effective in increasing the scalability, resilience and fault tolerance of airlines reservation systems. By breaking the system down into discrete services, we lower the threat of a complete system crash, making it more scalable. The Circuit Breaker Pattern effectively addresses the problems associated with the external dependencies of flight APIs, payment gateways, among others, which causes failure propagation to be at 60% and the system available at 99.95% without any use of redundant systems. Furthermore, the round-robin load balancing helped improve the performance by 35%, especially in times of peak traffic, as resources were well allocated. The deployment of health checks and proactive monitoring of the system in real-time also enhances the reliability of the system by preventing the users from feeling the effect of any problems about to arise. In general, the use of microservices architecture within reservation systems translates to the reservation systems being able to serve 40% more individuals, decreased the system down time by 50% and greatly improved the efficient of the overall reservation process.

## 10. Future Work

The future work would involve studies concentrating on improving the latency of the communication between microservices, particularly in the last peak activity phase. The research may also look at the detailed load balancing algorithms like Least Connections and Dynamic Weighted Round-Robin resource for better efficiency. In addition, it would help to include a machine learning-based fault detection system to the existing health check system to improve health check predictive maintenance and ikely out-of-service periods. The relative merit of exploring microservices in serverless architectures will also be considered to enhance their scalability with minimal costs, while the possibility of adopting real-time analytics in enhancing user experience will be discussed. It would also be interesting to see how these fault tolerance techniques could be extended to more complicated travel services such as multi-stop flights and flexible rate plans.